\documentclass{article}

\usepackage{arxiv}

\usepackage[utf8]{inputenc} 
\usepackage[T1]{fontenc}    
\usepackage{hyperref}       
\usepackage{url}            
\usepackage{booktabs}       
\usepackage{amsfonts}       
\usepackage{nicefrac}       
\usepackage{microtype}      
\usepackage{lipsum}
\usepackage{graphicx}
\usepackage{amssymb}
\usepackage{amsmath}
\usepackage{enumitem}
\usepackage{xcolor}
\usepackage{matlab-prettifier}
\usepackage[normalem]{ulem}
\usepackage[style=nature]{biblatex}
\usepackage{lineno}
\graphicspath{ {./images/} }
\hypersetup{hidelinks}

\title{Upper Bounds on Overshoot in SIR Models with Nonlinear Incidence}

\author{
 Maximilian M. Nguyen$^1$\\
  Lewis-Sigler Institute, Princeton University$^1$\\
  Carl Icahn Laboratory, Princeton, NJ 08544 \\
  \texttt{mmnguyen@princeton.edu}
}

\begin{document}
\maketitle
\pagestyle{plain}

\begin{abstract}
We expand the calculation of the upper bound on epidemic overshoot in SIR models to account for nonlinear incidence. We lay out the general procedure and restrictions to perform the calculation analytically for nonlinear functions in the number of susceptibles. We demonstrate the procedure by working through several examples and also numerically study what happens to the upper bound on overshoot when nonlinear incidence manifests in the form of epidemic dynamics over a contact network. We find that both steeper incidence terms and larger contact heterogeneity can increase the range of communicable diseases at which the overshoot remains a relatively large public health hazard.
\end{abstract}

\section*{Introduction}
Compartmental models have been an invaluable tool for analyzing the dynamics of epidemics for the last century. In particular, the susceptible-infected-recovered (SIR) model has long been a workhorse compartmental model for describing transient epidemics due to its relative simplicity and has received a lot of attention in both the academic literature and the public health arena \cite{anderson_infectious_1992, diekmann_mathematical_2013}. The mechanism of transmission of a communicable disease underlying this model and a variety of other contagion models is contact between healthy and infectious individuals, which results in conversion of the healthy individuals into an infected state. Choosing how to precisely define the transmission interaction between healthy and infectious individuals leads to a variety of models depending on the assumptions made. For instance, in a spatially-explicit model, one might represent the contact between individual members of the population as a network. Alternatively, if one is willing to assume all individuals within a compartment are identical, one can formulate a model using ordinary differential equations (ODEs) by assuming an incidence term for how the susceptible and infected compartments mix and generate new infected individuals.   

The quintessential SIR ODE model, known as the Kermack-McKendrick model, is shown in (\ref{eqn:S}-\ref{eqn:R}). The model assumes a bilinear incidence rate $\beta SI$ for the growth term of the infected compartment, with transmissibility parameter $\beta$ and first-order (i.e. linear) with respect to the fraction of population that is susceptible ($S$) and infected ($I$).

\begin{align}
\frac{dS}{dt}&=-\beta SI \label{eqn:S}\\
\frac{dI}{dt}&=\beta SI-\gamma I \label{eqn:I}\\
\frac{dR}{dt}&=\gamma I \label{eqn:R}
\end{align}

While a bilinear incidence rate between healthy and infected individuals can be a reasonable first assumption, depending on the real-world situation being modeled and the level of precision required, the assumption can be insufficient or inaccurate. A significant body of work in the literature has been done to generalize this incidence term into more detailed forms \cite{liu_influence_1986, liu_dynamical_1987, hethcote_epidemiological_1991, ruan_dynamical_2003, jin_sirs_2007, korobeinikov_global_2007}. Moving beyond bilinear forms towards nonlinear incidence allows for the consideration of models with more biological complexity and realism. Factors such as network effects, seasonality, and non-pharmaceutical interventions are known to give rise to more complex dynamics \cite{pastor-satorras_epidemic_2015, tkachenko_time-dependent_2021, gomes_individual_2022, montalban_herd_2022}. The studying of nonlinear transmission in the context of epidemiology here is a specific case of a larger growing interest across a range of fields in studying the effects of transmission dynamics in complex systems. The exploration of interactions beyond the bilinear form has taken place in contexts ranging from social dynamics \cite{axelrod_complexity_1997, centola_spread_2010, vespignani_modelling_2012,  lehmann_complex_2018, hebert-dufresne_macroscopic_2020}, ecology \cite{bascompte_rethinking_1995, hofbauer_evolutionary_1998, grilli_higher-order_2017, taylor_modified_2017}, economics \cite{may_ecology_2008, arthur_foundations_2021}, to molecular biology \cite{stefan_cooperative_2013}. Significant effort has also been expended on synthesizing these ideas into a more general model of contagion that can be used in different domains \cite{goffman_generalization_1964, barrat_dynamical_2008, centola_how_2018}. 

As the representation of transmission is arguably the most important aspect in setting up a model of communicable disease, the choice for the incidence rate has downstream consequences on many epidemiological quantities of interest. These include, but are not limited too, the epidemic size, the herd immunity threshold, epidemic duration, and the epidemic overshoot. While features such as the epidemic size and the herd immunity threshold have been studied rather extensively in the literature, the behavior of overshoot remains relatively under explored, in particular for more general incidence rates.

\begin{figure}[ht]
\centering
\includegraphics[scale=0.4]{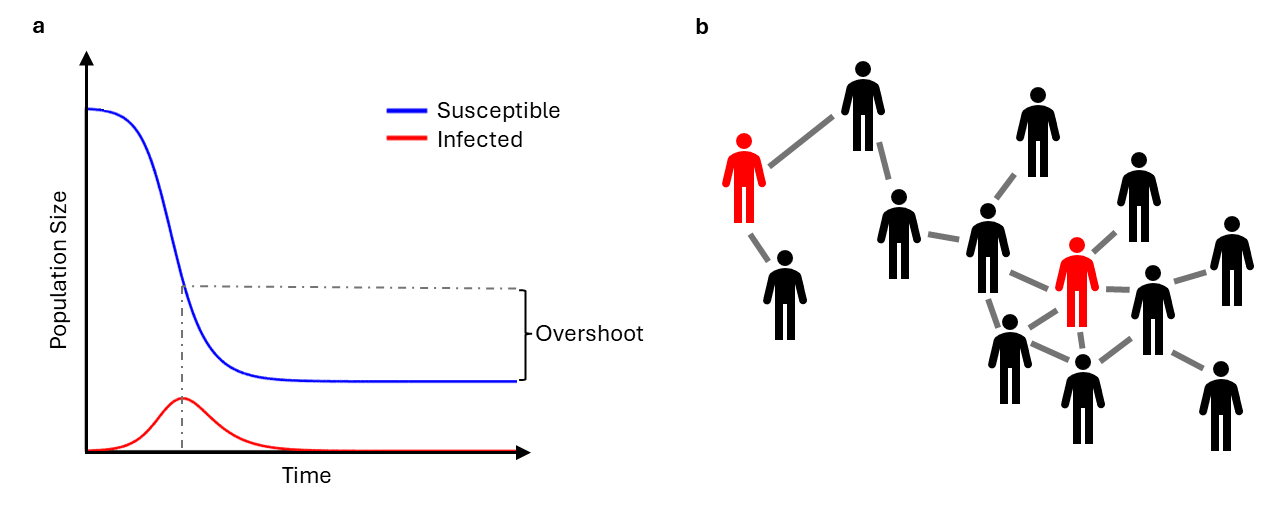}
\caption{Illustration of overshoot in the context of epidemics and spatial heterogeneity in contact networks. a) In the context of an epidemic outbreak, the overshoot is given by the depletion of susceptibles from the peak of the infections until the end of the outbreak. b) Nonlinear incidence can manifest in the form of a spatial proximity network.} \label{fig:cartoon}
\end{figure}

Overshoot is a concept from mathematics and control theory that quantifies the amount of excess from when a function exceeds its target value \cite{doyle_feedback_2013, sethi_what_2019}. This concept has been applied to a variety of contexts ranging from ecological and environmental problems that pertain to overconsumption and sustainability \cite{catton_overshoot_1982, wackernagel_tracking_2002, bradshaw_underestimating_2021, fanning_social_2022} to biotechnology that controls blood flow \cite{rosengarten_overshoot_2002}.  
In the epidemiological context, the overshoot quantifies the number of individuals that become infected after the prevalence peak of infections occurs (Figure \ref{fig:cartoon}a). In simple epidemics, as the peak of the epidemic coincides with the threshold at which transmission is sufficiently reduced so that the epidemic is no longer growing, the overshoot reflects the excess in cases beyond this minimal threshold of protection. 

While the terminology \textit{excess} may give the connotation of a relatively small effect, in the the Kermack-McKendrick SIR model it can be shown that up to nearly $30\%$ of the population can become infected in the overshoot phase of the epidemic \cite{nguyen_fundamental_2023}. This is also not a rare case, as large values of overshoot occur at rather common values for the basic reproduction ($R_0$) of $1.5-4$, which includes communicable diseases such as COVID-19 \cite{li_early_2020, darienzo_assessment_2020, majumder_early_2020}, HIV \cite{hollingsworth_hiv-1_2008}, and influenza \cite{biggerstaff_estimates_2014}. An analysis of data from the the first wave of the COVID-19 pandemic in the urban city of Manaus, Brazil \cite{buss_three-quarters_2021, nguyen_fundamental_2023}, where disease spread went largely unmitigated, suggested that the dynamics could be reasonably approximated by the Kermack-McKendrick SIR model and that nearly $30\%$ of the population became infected in the overshoot phase.

The overshoot highlights the potentially large public-health risk that can be posed by allowing for unmitigated spread or reducing intervention measures prematurely. However, the transmission rates at which the overshoot poses the greatest risk depends on the form of the incidence rate, which drives the need to understand the behavior for incidence beyond the simple bilinear case. As the overshoot quantifies the excess number of cases that occur after the herd immunity threshold has been reached, it is also intimately connected with any potential interventions or mitigation strategies. A question of great concern to epidemiologists and public health officials is figuring out the optimal control strategies for reducing excess cases and mortality \cite{handel_what_2006, lauro_optimal_2021, morris_optimal_2021, kollepara_ethical_2024}. Any optimal strategy by design generally seeks to eliminate the overshoot. Thus, a better understanding the behavior of overshoot under different model assumptions might allow for better control measures to be designed and developed.

While the overshoot within the SIR model has received some attention \cite{handel_what_2006,ellison_implications_2020, rachel_analytical_2020, ketcheson_optimal_2021}, a detailed understanding of its full mathematical behavior within compartmental models remains incomplete. An observation that the overshoot is largest for intermediate basic reproduction numbers was first numerically observed by Zarnitsyna et al. \cite{zarnitsyna_intermediate_2018}. An explanation for that phenomena in the context of the Kermack-McKendrick model was recently discovered, showing that the overshoot is derived from a trade-off from the basic reproduction number in driving both the final epidemic size and how quickly the disease burns through the population \cite{nguyen_fundamental_2023}. Here we lay the foundation to calculate the upper bound for overshoot when considering incidence terms beyond the simple bilinear case. We will first derive the behavior of the overshoot for more general incidence rates within the context of ODE models, where the results and understanding can be analytical and precise. As nonlinear incidence can also arise through the connectivity structure of networks (Figure \ref{fig:cartoon}b), we will then numerical explore the effect of network structure and changing network topologies on overshoot.

\section*{Results}
\subsection*{Effect of Nonlinear Incidence on Overshoot in SIR ODE Models}
We first examine the effect of nonlinear incidence on overshoot for ODE models, where the computations can be made analytical. The equations of the Kermack-McKendrick SIR ODE model are given as follows with generic incidence term $\beta f(S)g(I)$, where $f(S)$ and $g(I)$ are functions of $S$ and $I$ respectively to be specified.

\begin{align}
\frac{dS}{dt}&=-\beta f(S)g(I) \label{eqn:Sdot}\\
\frac{dI}{dt}&=\beta f(S)g(I)-\gamma I \label{eqn:Idot}\\
\frac{dR}{dt}&=\gamma I
\end{align}

where $S$, $I$, $R \in [0,1]$  are the fractions of the population that are susceptible, infected, and recovered respectively, $\beta,\gamma \in \mathbb R_{> 0}$ are positive-definite parameters for transmission and recovery rate respectively.

For the SIR model, the overshoot is given by the following equation:

\begin{align}
Overshoot = S_{t^*}-S_{\infty} \label{eqn:overshoot}
\end{align}

where $S_{t^*}$ is the fraction of susceptibles at the time of the prevalence peak (i.e. when $I$ is maximal in value), $t^*$, and $S_{\infty}$ is the fraction of susceptibles at the end of the epidemic. To solve this equation, the easiest approach is to derive an equation for $S_{t^*}$ in terms of only $S_{\infty}$ and parameters. We do this by first setting (\ref{eqn:Idot}) equal to 0 and solving for the critical susceptible fraction $S_{t^*}$. 

\begin{align}
\frac{dI}{dt} &= 0 = \beta f(S_{t^*}) g(I_{t^*}) - \gamma I_{t^*} \nonumber \\
\end{align}

By using the usual definition for the basic reproduction number, $R_0 \equiv \frac{\beta}{\gamma}$, we obtain the following equation for $S_{t^*}$.

\begin{align}
S_{t^*} &= f^{-1}\left(\frac{I_{t^*}}{g(I_{t^*})} \frac{1}{R_0}\right) \nonumber
\end{align}

We can see from this equation that $S_{t^*}$ will have $I$ dependence unless $g(I_{t^*})=I_{t^*}$. Thus to make what follows analytically tractable, let us assume $g(I_{t^*})=I_{t^*}$. We will provide even stronger justification why $g(I)$ must take this form later in the results. This assumption of $g(I_{t^*})=I_{t^*}$ reduces the above equation to the following.

\begin{align}
S_{t^*} &= f^{-1}\left(\frac{1}{R_0}
\right) \label{eqn:Sstar}
\end{align}
Taking this equation for $S_{t^*}$ (\ref{eqn:Sstar}) and the overshoot formula (\ref{eqn:overshoot}), we obtain: 

\begin{align}
Overshoot = f^{-1}\left(\frac{1}{R_0}
\right)-S_{\infty} \label{eqn:overshoot_new}
\end{align}

Thus the main challenge now becomes a problem of finding an equation for $R_0$ and the inverse function $f^{-1}$. Based on previous results \cite{nguyen_fundamental_2023}, the following outlines the general steps for calculating the maximal overshoot for a SIR model:

\begin{enumerate}[label=\Alph*.]
    \item Take the ratio of $\frac{dI}{dt}$ and $\frac{dS}{dt}$. Integrate the resulting ratio. The indefinite integral requires a constant of integration, which is a conserved quantity that applies at every time point along the system's trajectory in time. \label{step1}
    \item Evaluate the equation for the conserved quantity at the beginning of the epidemic ($t=0$) and the end of the epidemic ($t=\infty$) using initial conditions and asymptotic values. Then, rearrange the resulting equation for $\frac{1}{R_0}$. \label{step2}
    \item Find the form for the inverse function, $f^{-1}$. \label{step3}
    \item Combine the equations for $\frac{1}{R_0}$ and $f^{-1}$ with the overshoot equation. \label{step4}
    \item Maximize the resulting overshoot equation by taking the derivative of the equation with respect to $S_{\infty}$ and setting the equation to 0 to find the extremal point $S_{\infty}^*$. This step usually leads to a transcendental equation for $S_{\infty}^*$, which can be solved numerically. \label{step5}
    \item Use the maximizing $S_{\infty}^*$ value in the overshoot equation to calculate the corresponding maximal overshoot. \label{step6}
    \item Calculate the corresponding $R_0^*$ using $S_{\infty}^*$ and the $\frac{1}{R_0}$ equation. \label{step7}
\end{enumerate}

Thus, the analytical exploration of nonlinear incidence terms of the type $\beta f(S)g(I)$ is reduced to exploring different forms of $f(S)$.

\subsection*{Restrictions on $g(I)$}
The first step is to rule out what forms for the incidence term will not work with the procedure outlined above.

We now show the principle reason why we require $g(I)=I$. We can see from calculating Step \ref{step1} in (\ref{eqn:ruleoutI}) that any incidence term that does not take the form $g(I)=aI, a \in \mathbb{R}$, where $a$ is a real scalar, retains $I$ dependence upon simplification. 

\begin{align}
\text{Step \ref{step1}: }\frac{\frac{dI}{dt}}{\frac{dS}{dt}} = \frac{\beta f(S)g(I)-\gamma I}{-\beta f(S)g(I)} = -1 + \frac{I}{R_0 f(S) g(I)} \label{eqn:ruleoutI}
\end{align}

Any deviation from the form $g(I)=aI$ results in $I$ in the numerator and the denominator not completely cancelling out which will result in having to integrate $I$ with respect to $S$, which we will not be able to do analytically. Therefore, $I$ must enter linearly into the incidence term. Since $a$ can be absorbed into the $\beta$ parameter, all possible incidence terms for the purpose of calculating overshoot analytically will take the form $\beta \cdot f(S) \cdot I$. 

\subsection*{Restrictions on $f(S)$}
We now turn to what restrictions there are on the form of $f(S)$. We start first with two conditions. First, we must enforce that when there are no susceptibles ($S=0$), then the incidence rate must go to zero ($\beta f(S)I=0$). Otherwise, since $I$ does not have such a restriction, violating this condition would leave open the unrealistic possibility that the model can generate infected people when there are no susceptibles available. To ensure this condition is met, we need the function on S to output zero if the input is zero.

\begin{align}
f(S=0)=0 \label{eqn:BC1}
\end{align}

For the second condition, for an outbreak to occur in an SIR model, we must have a minimum value for the basic reproduction number, $R_0$. Another way to view this condition is by inspecting (\ref{eqn:I}) and recognizing that that the incidence term ($\beta f(S)I$) needs to be greater than the recovery term ($\gamma I$). Otherwise the epidemic cannot grow in size. Comparing the two terms leaves an inequality ($\beta f(S)I > \gamma I$) which can be rewritten as:

\begin{align}
R_0 > \frac{1}{f(S)} \label{eqn:BC2}
\end{align}

Beyond these conditions, an obvious requirement is that $f(S)$ should be a continuous function. In order to be able to calculate the maximal overshoot analytically, the function should be integratable with respect to $S$ and should also have a closed-form inverse $f^{-1}$. As we will demonstrate, non-monotonic functions for $f(S)$ are possible.

For $f(S)$, the following examples are constructed using basic functions that satisfy the above criteria:
\begin{enumerate}[nosep]
    \item $\text{exp}(S) - 1$
    \item Invertible polynomials of $S$
    \item $\text{sin}(aS)$
\end{enumerate}

Conversely, there are many examples of functions that would not work. An example that satisfies the boundary conditions but that does not have a closed form inverse is $f(S)=\text{log}(S+1)$. While similar to examples listed, the following violate the boundary conditions: $\text{exp}(S), \text{log}(S)$, $\text{cos}(S)$. Examples that violate conditions of continuity include step functions of $S$ or $f(S)$ with cusps.

\subsection*{Deriving Maximal Overshoot for Various $f(S)$}

As an example, we will now look at a model with nonlinear incidence and apply the whole procedure previously outlined for finding the maximal overshoot.

\subsubsection*{Example: $f(S)=\text{exp}(S)-1$}
Let us consider an incidence rate that takes on an exponential form. This produces an incidence term that grows slightly faster than the original bilinear incidence term, and so might be relevant in situations where there are network effects.The phenomenon of superspreading occurs when some individuals infect many more people than the typical infected person would normally do \cite{majra_sars-cov-2_2021}. Superspreading allows for explosive outbreaks that exceed the growth rate of infections allowed by simple bilinear incidence and is enabled by inherent heterogeneity in contacts amongst the population, which makes it natural to occur in situations that are approximated by a network. Different models can be formulated to account for superspreading \cite{fujie_effects_2007, nielsen_covid-19_2021}. Here we have chosen the exponential as a qualitative simple example that more closely resembles the growth rate on a heterogeneous network than standard bilinear incidence. 

We start at Step \ref{step1} by solving for the rate of change of $I$ as a function of $S$ by taking the ratio of $\frac{dI}{dt}$ and $\frac{dS}{dt}$.

\begin{align}
\frac{dI}{dS} &= -1 + \frac{1}{R_0 (e^S-1)} \label{eqn:exp} 
\end{align}
from which it follows on integration using the substitution $u=e^S$ and partial fractions $\left(\frac{1}{u-1}\right)$ and $\left(\frac{1}{u}\right)$ that $I + S + \frac{S-\text{ln}|e^S-1|}{R_0}$ is constant along all trajectories.

For Step \ref{step2}, consider the conserved quantity at both the beginning ($t=0$) and end ($t = \infty$) of the epidemic. 

\begin{align}
I_0 + S_0 + \frac{S_0-\text{ln}|e^{S_0}-1|}{R_0} &= I_{\infty} + S_{\infty} + \frac{S_{\infty}-\text{ln}|e^{S_{\infty}}-1|}{R_0}  \nonumber
\end{align}
hence

\begin{align}
\frac{1}{R_0} = \frac{I_{\infty}+S_{\infty}-I_0-S_0}{S_0-S_{\infty}+\text{ln}\left(\frac{|e^{S_{\infty}}-1|}{||e^{S_0}-1|}\right)}
\end{align}

We use the initial conditions: $S_0=1-\epsilon$ and $I_0=\epsilon$, where $\epsilon$ is the (infinitesimally small) fraction of initially infected individuals. We use the standard asymptotic of the SIR model that there are no infected individuals at the end of an SIR epidemic: $I_{\infty} = 0$. This yields:

\begin{align}
\frac{1}{R_0} &= \frac{S_{\infty} -1}{1- S_{\infty}+\text{ln}\left(\frac{|e^{S_{\infty}}-1|}{e-1}\right)} \label{eqn:R0_2}
\end{align}

For Step \ref{step3}, we find the inverse of $f$. 

\begin{align}
f(x) = e^S-1 \implies f^{-1}(x)=\text{ln}(x+1) \label{eqn:inverse_2}
\end{align}

For Step \ref{step4}, we substitute the expression for $\frac{1}{R_0}$ (\ref{eqn:R0_2}) and $f^{-1}$ (\ref{eqn:inverse_2}) into the overshoot equation (\ref{eqn:overshoot_new}).

\begin{align}
Overshoot &= \text{ln}\left(\frac{\text{ln}\left(\frac{|e^{S_{\infty}}-1|}{e-1}\right)}{1- S_{\infty}+\text{ln}\left(\frac{|e^{S_{\infty}}-1|}{e-1}\right)}\right)- S_{\infty} \label{eqn:overshoot_final2}
\end{align}

For Step \ref{step5}, differentiation of both sides with respect to $S_{\infty}$ and setting the equation to zero to solve for the critical $S_{\infty}^*$ yields:

{\tiny
\begin{align}
0 &= \left(\frac{\text{ln}\left(\frac{|e^{S^*_{\infty}}-1|}{e-1}\right)}{1- S^*_{\infty}+\text{ln}\left(\frac{|e^{S^*_{\infty}}-1|}{e-1}\right)}\right)^{-1}\left(\frac{\left(1-S^*_{\infty}+\text{ln}\left(\frac{|e^{S^*_{\infty}}-1|}{e-1}\right)\right)\cdot\left(\left(\frac{|e^{S^*_{\infty}}-1|}{e-1}\right)^{-1}\frac{e^{S^*_{\infty}}}{e-1}\right) - \text{ln}\left(\frac{|e^{S^*_{\infty}}-1|}{e-1}\right))\cdot\left(-1 + \left(\frac{|e^{S^*_{\infty}}-1|}{e-1}\right)^{-1}\frac{e^{S^*_{\infty}}}{e-1}\right) }{\left(1-S^*_{\infty}+\text{ln}\left(\frac{|e^{S^*_{\infty}}-1|}{e-1}\right)\right)^2}\right) - 1 \nonumber
\end{align}}%
Since $e^S-1$ is positive semi-definite over the unit interval for $S$, dropping the absolute value symbols and simplifying yields:

\begin{align}
\text{ln}\left(\frac{e^{S^*_{\infty}}-1}{e-1}\right)\left(\text{ln}\left(\frac{e^{S^*_{\infty}}-1}{e-1}\right)- S^*_{\infty}\right) &= (1-S^*_{\infty})\left(\frac{e^{S^*_{\infty}}}{e^{S^*_{\infty}}-1}\right) \nonumber \label{eqn:trans_2}
\end{align}
which admits both a trivial solution ($S_{\infty}^* = 1$) and the solution $S_{\infty}^* = 0.1663...$ .

For Step \ref{step6}, using the non-trivial solution for $S_{\infty}^*$ in the overshoot equation (\ref{eqn:overshoot_final2}) to obtain the value of the maximal overshoot for this model, $Overshoot^* |_{\substack {_{\beta (e^S-1)I}}}$ yields:

\begin{align}
Overshoot^*|_{\substack {_{\beta (e^S-1)I}}} &= 0.2963...
\end{align}

Thus, the maximal overshoot for incidence functions of the form $\beta (e^S-1)I$ is $0.296...$ .

\begin{figure}[ht]
\centering
\includegraphics[scale=0.28]{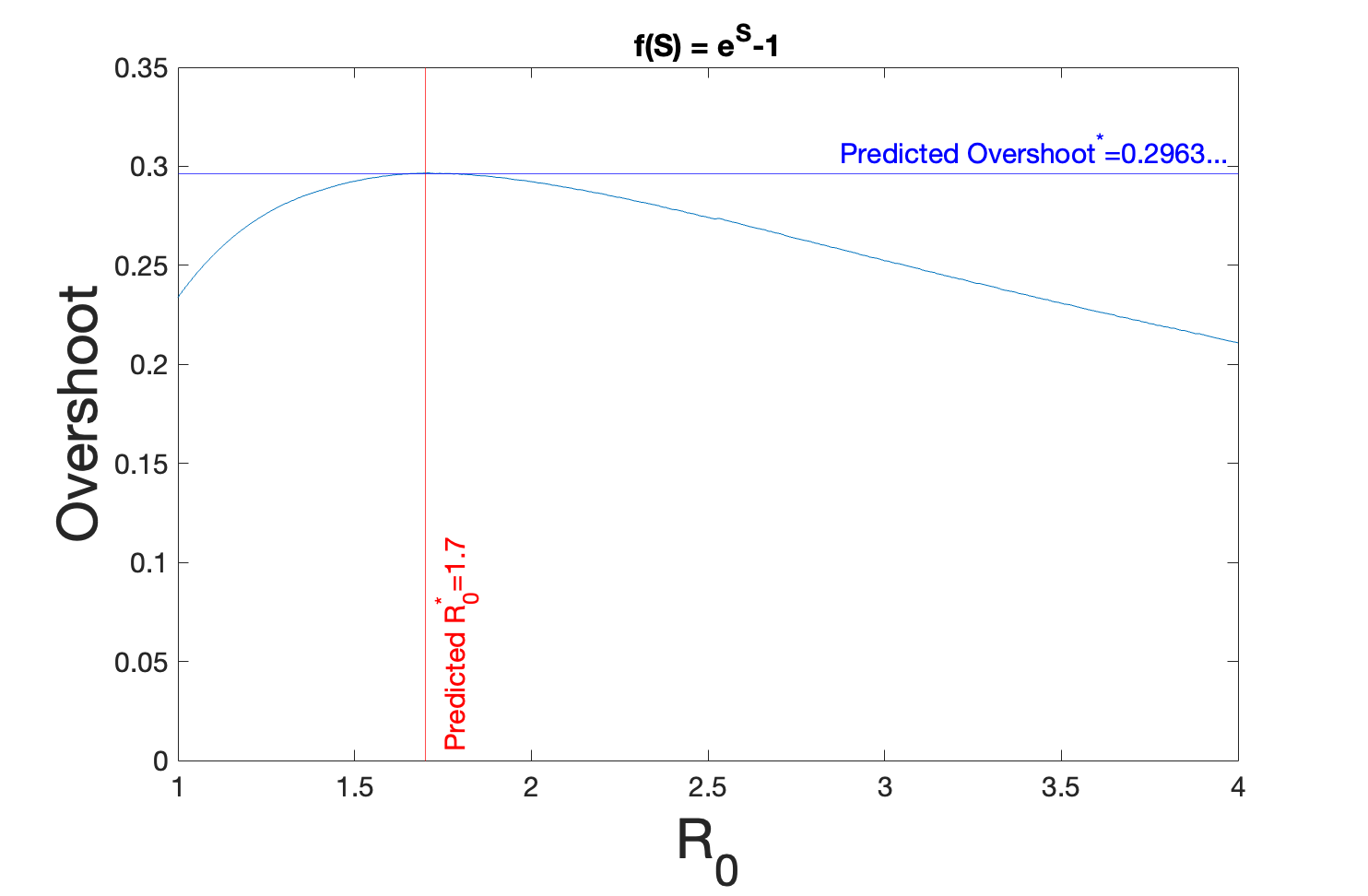}
\caption{The overshoot in a model with exponential incidence. The overshoot as a function of $R_0$ for an SIR model with nonlinear incidence term of $\beta (e^S-1) I$. The horizontal line for $Overshoot^*$ in dark blue and the vertical line in red given by $R_0^*=1.7$ are the theoretical predictions given by the calculations in the text. The curve is obtained from numerical simulations.} \label{fig:overshoot_expo}
\end{figure}

For Step \ref{step7}, we can calculate the corresponding $R_0^*$ using $S_{\infty}^*$ and (\ref{eqn:R0_2}).

\begin{align}
R_0^*|_{\substack {_{\beta (e^S-1)I}}} &= 1.7
\end{align}
This result is verified numerically in Figure \ref{fig:overshoot_expo}. Compared to the overshoot in the Kermack-McKendrick model \cite{zarnitsyna_intermediate_2018, nguyen_fundamental_2023}, we see that the maximal overshoot is also near $30\%$. However, we see that overshoot has a much broader distribution over the domain of $R_0$, suggesting that exponential incidence rates pose a public health hazard for a greater range of communicable diseases over their bilinear counterparts. 

Additional examples for incidence rates of a different mathematical form can be found in the Supplemental Information. One example considers when the form of $f(S)$ is given by a polynomial, which can be used to model higher-order effects beyond a simple two-person interaction. Another example considers the consequence of having $f(S)$ that is non-monotonic over the domain of $S$, which might reflect real scenarios where there are tipping points in behavior.

\subsection*{Nonlinear Incidence Generated from Dynamics on Networks}
In the previous sections, we focused on nonlinear incidence that was generated from the introduction of a nonlinearity in an ordinary differential equation model. Importantly, this fundamentally assumes homogeneity in the transmission, in that all infected individuals are identical in their ability to spread the disease further. In contrast, network models allow for heterogeneous spreading which depends on the local connectivity of each infected individual. This provides an entirely different mechanism through which nonlinear incidence can be generated compared to the ordinary differential equation models. As many real-world complexities and details can be more easily captured and explored in a network model, the consequences of those heterogeneities on the overshoot becomes more transparent through the targeted and fully-fleshed out explorations that can be done through numerical experimentation in network simulations \cite{newman_spread_2002, wang_epidemic_2003, keeling_networks_2005, pastor-satorras_epidemic_2015, kiss_mathematics_2017}. However, a trade-off for the increased realism is that it becomes more difficult to perform analytical calculations for a general network model, so here we conduct a numerical exploration of the behavior of the overshoot in network models across a range of network structure.

The space of all possible network configurations is immense, so we must restrict ourselves to analyzing a particular subset of possible networks. Here we explore what happens to the overshoot when the contact structure of the population is given by a network graph that is roughly one giant component. While it is possible to construct pathological graphs that produce very complex dynamics, we consider more classical graphs here. Using a parameterization of heterogeneity ($\sigma$) given by Ozbay et al. \cite{ozbay_parameterizing_2023} and the configuration model of Newman \cite{newman_networks_2018} to randomly construct networks (see Methods for details), we simulated epidemics on a spectrum of networks with structure ranging from the homogeneous limit (well-mixed, complete graph) to a heterogeneous limit (heavy-tailed degree distributions). The spectrum of distributions shapes across the space of $\sigma$ used to generate the degree distributions can be seen in the supplement. As the dynamics of the epidemic on a network are stochastic and occur in discrete-time here, they are not parameterized by $R_0$ as in the ODE models. Instead, we use an analogous parameter that we will denote as a basic reproduction number for networks ($R_{0,network}$) and is defined as follows:
\begin{equation}
R_{0,network} \equiv \frac{\tau \cdot \langle k \rangle} {\rho}
\end{equation}
The transmission probability ($\tau$) and recovery probability ($\rho$) are directly analogous to their counterparts ($\beta$ and $\gamma$ respectively) in $R_0$ in that that correspond to transmission and recovery parameters, albeit for a stochastic model. The inclusion of the mean network degree ($\langle k \rangle$) as a scalar makes intuitive sense as that represents the average number of potential neighbors an infected node could potentially infect at the beginning of the epidemic, which is analogous to the interpretation of $R_0$ as the number of secondary infections. We observed what happens to the overshoot on these different graphs as we changed $R_{0,network}$. 

\begin{figure}
\centering
\includegraphics[scale=0.6]{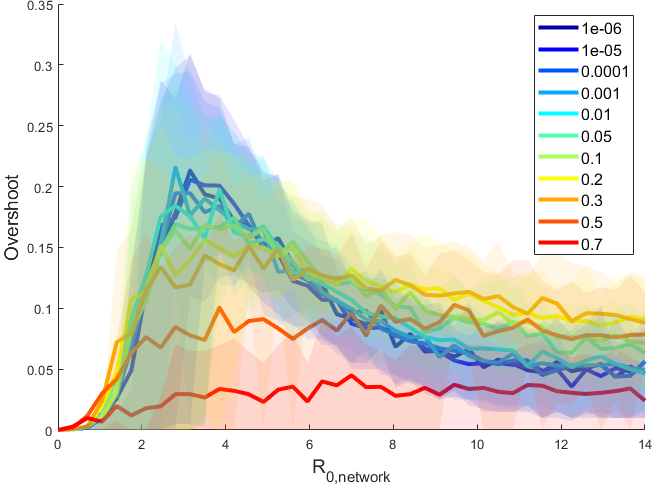}
\caption{Comparing the overshoot in the context of networks with different amounts of contact heterogeneity. The overshoot for SIR epidemic simulations on networks with varying levels of heterogeneity ($\sigma$) as a function of $R_{0, network}$. Each color shows simulations for networks of different contact heterogeneity as parameterized by $\sigma$. The solid lines represent the mean value of 100 simulation runs for a given $R_{0,network}$ and $\sigma$. The shaded areas indicate the 25th and 75th quartiles for those 100 simulations. The other simulation parameters are number of nodes (N) = 200, $\langle k \rangle$ = 7, and recovery probability ($\rho$) = 0.2.} \label{fig:hetero}
\end{figure}

On a network model, where contact structure is made explicit, the homogeneous limit is a complete graph, which recapitulates the well-mixed assumption of the Kermack-McKendrick model. It is not surprising then that the overshoot in the homogeneous graphs ($\sigma \approx 0)$ peaks also around 0.3 (Figure \ref{fig:hetero}, $purple$ and $blue$), which coincides with the analytical upper bound previously found in the ordinary differential equation Kermack-McKendrick model \cite{nguyen_fundamental_2023}. However, while the homogeneous graphs shown here are highly regular (i.e. symmetric in the average number of contacts each node has), the network's connectivity is not close to complete as the mean degree is significantly less than the network size. Thus, the average overshoot is lower than would be the case for a complete graph.

We also see that increasing contact heterogeneity (i.e. $\sigma \rightarrow 1$) qualitatively suppresses the overshoot peak both in terms of the overshoot value and the corresponding $R_{0,network}$. Furthermore, increased heterogeneity also flattens out the overshoot curve as a function of $R_{0,network}$. We see that for more heterogeneous graphs, the overshoot is larger at very low $R_{0,network}$. For some intermediate values of contact heterogeneity (Figure \ref{fig:hetero}, $yellow$ and $orange$), the overshoot shows larger overshoot than the homogeneous case for large $R_{0,network}$. This would suggest a larger public health hazard for a larger range of communicable diseases for networks with structure in this regime. The intuition for the larger overshoot at high $R_{0,network}$ in more heterogeneous networks is that both the peak occurs earlier and that a significant number of cases can occur in the periphery of a network (Supplemental Materials). The probability of being connected to a high-degree node increases as the heterogeneity of the network increases, which drives the peak of infections to occur sooner. In the second phase of the epidemic as the epidemic expands out to the periphery the epidemic burns more slowly as individuals in the periphery have fewer neighbors.

However, this trend of a overshoot distribution with a big right tail does not monotonically increase with the contact heterogeneity. We see that at very high amounts of contact heterogeneity, the overall area under this overshoot curve decreases. This can be partially explained by the fact that for very heterogeneous networks, a significant fraction of the population have no neighbors at all. This limits the amount of people that can potentially be infected and caps the outbreak size and subsequently overshoot size.

\section*{Discussion}
 While the overshoot has received less attention and exploration than its epidemiological counterpart (the herd immunity threshold), the overshoot poses a significant potential public health hazard for a large range of communicable diseases when there is no mitigation. Generalizing models to include nonlinear incidence terms allows for the consideration of more real-world effects such as higher-order transmission effects and network effects. Thus, to expand the scope of previous work, we have illustrated a general method to analytically find the maximal overshoot for generic nonlinear incidence terms.

Starting with the general incidence term $\beta f(S)g(I)$, we have deduced what restrictions must be placed on the form of $f(S)$ and $g(I)$ to make an analytical calculation possible. As long as the conditions for a suitable $f(S)$ are satisfied, in principle the maximal overshoot can be derived. However, in the examples shown  in both the main text and supplemental information, we have seen that even relatively simple forms for $f(S)$ can quickly lead to complicated integrals and derivatives. For these examples, we have shown the predictions given by the theoretical calculations generally match the empirical results derived from numerical simulation. From a public-health perspective, we find that incidence rates that are steeper over the domain of $S$, such as the examples with the exponential or polynomial functions, show a smaller maximal overshoot, but importantly, a much broader range of $R_0$'s at which the the overshoot remains large. As interventions that seek optimal control try to minimize the overshoot, this highlights the need for even stronger interventions when the interactions and subsequent incidence between individuals is higher than a simple bilinear interaction. The example of $f(S)=\text{sin}(aS)$ in the supplemental information is interesting because it can be used to probe the restrictions on the shape of $f(S)$. The case demonstrated that $f(S)$ no longer has to be monotonic over the domain of $S$, allowing for tipping point-like behavior.

Nonlinear incidence introduced through having a network structure showed that having network connectivity that was more heterogeneous resulted in a reduction in the upper bound on overshoot and a reduction of the dependence of overshoot on transmission overall. Numerically, the overshoot for a homogeneously connected network (i.e. a complete graph) well-approximates a Kermack-McKendrick ODE model. The ODE model makes a fundamental assumption of a well-mixed population, and similarly, a network with the highest mixing rate would be a complete graph, which a graph approaches as it becomes more homogeneous and the connectivity increases. Thus, an upper bound on the overshoot of approximately $0.3$ in both models \cite{nguyen_fundamental_2023} is perhaps unsurprising. It is interesting to note though, that due to the stochastic nature of the network simulations, occasionally an epidemic on a very homogeneous contact network will exceed the analytical bound in the ODE framework (see Figure \ref{fig:hetero}, the upper quartiles for $blue$ and $purple$ regions exceed $30\%$). While more heterogeneous networks did not have an overshoot larger than $0.3$, the overshoot was greater at larger $R_{0,network}$ than the homogeneous case. Future studies that incorporate more nuanced aspects of behavior, interventions, and time dynamics in the network model can further elucidate the complexities of epidemics on networks.

It will also be interesting to see if more complicated nonlinear interaction terms than the ones presented here can be derived. In addition, it will also be interesting to see how these nonlinear incidence terms interact when additional complexity is added to the SIR model, such as the addition of vaccinations or multiple subpopulations.

\section*{Methods}
\subsection*{Generating Graphs of Differing Heterogeneity}
In Figure \ref{fig:hetero}, we presented the results of SIR simulations of epidemics run on graphs of size N = 200 and the mean degree is 7, where the parameter of interest is $\sigma$. Each curve represents a different value of the graph heterogeneity. We implemented the following procedure from \cite{ozbay_parameterizing_2023} for generating graphs as a function of a continuous parameter ($\sigma$):

The following simple procedure generates a graph that has the desired heterogeneity: 
\begin{enumerate}
     \item Choose values of $\sigma$ (heterogeneity), the mean node degree (which can be set through the relationship $\text{Mean} = \lambda \Gamma \left(1-\frac{1}{\text{log}(\sigma)}\right)$ via $\sigma$ and an appropriate scale parameter ($\lambda$), and $N$ (number of nodes).
     \item Draw N random samples from the following distribution using $\sigma$ and $\lambda$, rounding these samples to the nearest integer, since the degree of a node can only take on integer values.
     $$f(x; \lambda, \sigma) = \frac{-\text{\text{ln}} (\sigma)}{\lambda}\left(\frac{x}{\lambda}\right)^{-\text{\text{ln}} (\sigma) - 1} e^{-(x/\lambda)^{-\text{\text{ln}} (\sigma)}};   x\ge 0; \sigma \in (0,1] ; \lambda > \mathbb{R}^+$$
    
     \item With the sampled degree distribution from the previous step, now use the configuration model method \cite{newman_networks_2018} (which samples over the space of all possible graphs corresponding to a particular degree distribution) to generate a corresponding graph.
\end{enumerate}  
This yields a valid graph with the desired amount of heterogeneity as specified by $\sigma$.

\subsection*{Simulating Epidemics on Graphs of Differing Heterogeneity}
We implemented the following simulation procedure from \cite{ozbay_parameterizing_2023} for implementing an SIR epidemic on a graph:

Given a graph $G(\sigma)$ of heterogeneity $\sigma$, fix a transmission probability $\tau$ and recovery probability $\rho$:

\begin{enumerate}
\item At time $t_0$, fix a small fraction $f$ of nodes to be chosen uniformly on the graph and assign them to the Infected state. The remaining $(1 - f)$ fraction of nodes start as Susceptible. 
\item For each $i \in [1,T]$, for each pair of adjacent S and I nodes, the susceptible node becomes infected with probability $\tau$.
\item For each $i \in [1,T]$, each infected node recovers with probability $\rho$.
\item For each simulation, record the overshoot as the difference between the number of susceptibles at the peak of infection prevalence and the end of the time dynamics.
\item Repeat steps (1-4) $n$ times for each value of $\tau$.
\item Repeat steps (1-5) for each value of $\sigma$.
\end{enumerate}

Code for the network simulations is provided in the Supplemental Information.

\subsection*{Numerical Solutions and Code}
Where needed, equations were solved numerically using the $ode45$ numerical solver in $MATLAB$. Code for the network simulations is provided in the Supplemental Information.

\section*{Acknowledgements}
The author would like to acknowledge generous funding support provided by the National Science Foundation (DMS-2327711) and a gift from the William H. Miller III.

\section*{Author Contributions}
M.M.N. designed research, performed research, and wrote and reviewed the manuscript.

\section*{Competing Interests}
The author declares no competing financial or non-financial interests.

\section*{Data Availability Statement}
Data sharing is not applicable to this article as no new data were created or analyzed in this study.

\printbibliography




\newpage
\begin{center}
\textbf{\large Supplemental Materials: Upper Bounds on Overshoot in SIR Models with Nonlinear Incidence}
\end{center}
\setcounter{figure}{0}
\renewcommand{\figurename}{Supplemental Figure}
\renewcommand{\thefigure}{\arabic{figure}}
\setcounter{equation}{0}
\pagestyle{plain}

\subsubsection*{Example 2: $f(S)$ = Invertible Polynomials of $S$}
Next, let us consider invertible polynomials. Having access to large degree polynomials will allow us to make arbitrarily sharp incidence terms. This allows one to make the growth rate of the epidemic even sharper than the exponential case if needed. Or in general, having access to the toolbox of all polynomials will allow one to pick an appropriately steep incidence term that best matches the data of the particular outbreak being analyzed. While the set of invertible polynomials is a relatively small subset of all polynomials, it still contains an infinitely large number of possible functions. Because the procedure will require integration of $f(S)$ while it is in the denominator, the algebraic details of doing this for higher-order polynomials with lower-order terms can quickly become cumbersome. So let us illustrate a test case using just the leading term of a generic cubic function. Let $f(S) = aS^3$, where $a \in \mathbb{R}$.

We start at Step \ref{step1} by solving for the rate of change of $I$ as a function of $S$ by taking the ratio of $\frac{dI}{dt}$ and $\frac{dS}{dt}$.

\begin{align}
\frac{dI}{dS} &= -1 + \frac{1}{R_0 aS^3} \label{eqn:cubic}
\end{align}
from which it follows on integration that $I + S + \frac{1}{2R_0 aS^2}$ is constant along any trajectory.

For Step \ref{step2}, consider the conserved quantity at both the beginning ($t=0$) and end ($t = \infty$) of the epidemic.

\begin{align}
I_0 + S_0 + \frac{1}{2R_0 aS_0^2} &= I_{\infty} + S_{\infty} + \frac{1}{2R_0 aS_{\infty}^2} \nonumber
\end{align}
hence

\begin{align}
\frac{1}{R_0} &= (I_{\infty} + S_{\infty} - I_0 - S_0)\frac{2a(S_0 S_{\infty})^2}{S_{\infty}^2-S_0^2} 
\end{align}

Using the initial conditions ($S_0=1-\epsilon$ and $I_0=\epsilon$, where $\epsilon << 1$) and asymptotic condition ($I_{\infty}=0$) yields:

\begin{align}
\frac{1}{R_0} &= \frac{2aS_{\infty}^2}{S_{\infty}+1} \label{eqn:R0_1}
\end{align}

For Step \ref{step3}, we find the inverse of $f$.

\begin{align}
f(x) = ax^3 \implies f^{-1}(x)=\left(\frac{x}{a}\right)^{1/3} \label{eqn:inverse_1}
\end{align}

For Step \ref{step4}, we substitute the expression for $\frac{1}{R_0}$ (\ref{eqn:R0_1}) and $f^{-1}$ (\ref{eqn:inverse_1}) into the overshoot equation ($Overshoot = f^{-1}\left(\frac{1}{R_0}
\right)-S_{\infty}$).

\begin{align}
Overshoot &= \left(\frac{2S_{\infty}^2}{S_{\infty}+1}\right)^{1/3}- S_{\infty} \label{eqn:overshoot_final1}
\end{align}

For Step \ref{step5}, differentiation with respect to $S_{\infty}$ and setting the equation to zero to solve for the critical $S_{\infty}^*$ yields:

\begin{align}
0 &= \frac{1}{3} \left(\frac{2S_{\infty}^{*2}}{S_{\infty}^*+1}\right)^{-2/3}\left(\frac{(S_{\infty}^*+1)\cdot 4S_{\infty}^* - 2S_{\infty}^{*2}\cdot1}{(S_{\infty}^*+1)^2}\right) - 1
\end{align}
whose solution is $S_{\infty}^* = 0.310...$ .

For Step \ref{step6}, using $S_{\infty}^*$ in the overshoot equation (\ref{eqn:overshoot_final1}) to obtain the value of the maximal overshoot for this model, $Overshoot^* |_{\substack {_{\beta (aS^3)I}}}$ yields:

\begin{align}
Overshoot^*|_{\substack {_{\beta (aS^3)I}}} &= 0.217...
\end{align}
Thus, the maximal overshoot for incidence functions of the form $\beta (aS^3)I$ is 0.217... .

For Step \ref{step7}, we can calculate the corresponding $R_0^*$ using $S_{\infty}^*$ and (\ref{eqn:R0_1}).

\begin{align}
R_0^*|_{\substack {_{\beta (aS^3)I}}} = \frac{6.816...}{a}
\end{align}
This prediction of the maximal overshoot being independent of $a$, whereas the corresponding critical $R_0$ is inversely proportional to $a$ is verified numerical in Supplemental Figure \ref{fig:overshoot_cubic}. Furthermore, it can be shown that the independence of the maximal overshoot of $a$ carries holds for any $f(S)=aS^n$. Since $a$ can be absorbed into $\beta$, the maximal overshoot is consequently independent of $R_0$ and instead only depends on the power of $S$.

\begin{figure}[ht]
\centering
\includegraphics[scale=0.3]{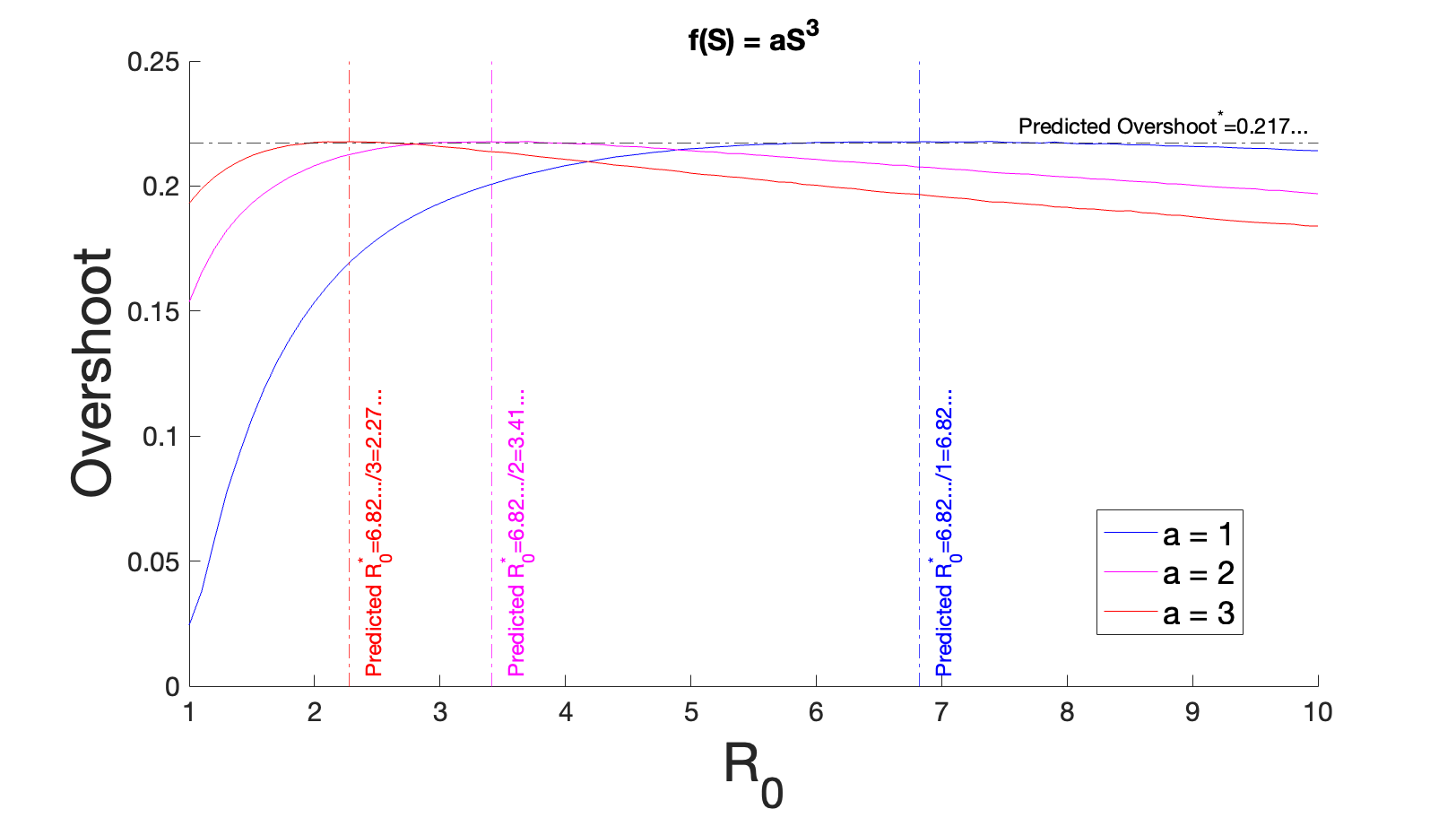}
\caption{The overshoot as a function of $R_0$ for an SIR model with nonlinear incidence term of $\beta (aS^3) I$ for different values of $a$. The dashed horizontal line for $Overshoot^*$ and the dashed vertical lines given by $R_0^*=1/a$ are the theoretical predictions given by the calculations in the text. The solid curves are obtained from numerical simulations using the value of the $a$ parameter.} \label{fig:overshoot_cubic}
\end{figure}

\subsubsection*{Example 3: $f(S)=\text{sin}(aS), a>0$}
The last example considered here is a periodic function of $S$, namely $sin(aS)$. This might be helpful in a scenario where one gets periodic fluctuations in the incidence rate. This might be relevant for scenarios that try to capture cyclical effects such as seasonality \cite{grassly_seasonal_2006}. Some models have accounted for seasonality in the incidence through incorporating a periodic function into the transmission parameter \cite{altizer_seasonality_2006, aron_seasonality_1984}. Here we make the assumption the periodicity in the incidence enters through the susceptibles.

We start at Step \ref{step1} by solving for the rate of change of $I$ as a function of $S$ by taking the ratio of $\frac{dI}{dt}$ and $\frac{dS}{dt}$.

\begin{align}
\frac{dI}{dS} &= -1 + \frac{1}{R_0 \text{sin}(aS)} \nonumber
\label{eqn:sine}
\end{align}
from which it follows on integration using the substitution $u=\text{csc}(aS)+\text{cot}(aS)$ that $I + S + \frac{\text{ln}|\text{csc}(aS)+\text{cot}(aS)|}{R_0a}$ is constant along all trajectories.

For Step \ref{step2}, consider the conserved quantity at both the beginning ($t=0$) and end ($t = \infty$) of the epidemic.

\begin{align}
I_0 + S_0 + \frac{\text{ln}|\text{csc}(aS_0)+\text{cot}(aS_0)|}{R_0a} &= I_{\infty} + S_{\infty} + \frac{\text{ln}|\text{csc}(aS_{\infty})+\text{cot}(aS_{\infty})|}{R_0a} \nonumber 
\end{align}
hence

\begin{align}
\frac{1}{R_0} = (I_{\infty}+S_{\infty}-I_0-S_0) \frac{a}{\text{ln}\frac{|\text{csc}(aS_0)+\text{cot}(aS_0)|}{|\text{csc}(aS_{\infty})+\text{cot}(aS_{\infty})|}}
\end{align}

Using the initial conditions ($S_0=1-\epsilon$ and $I_0=\epsilon$, where $\epsilon << 1$) and asymptotic condition ($I_{\infty}=0$) yields:

\begin{align}
\frac{1}{R_0} &= \frac{a(S_{\infty} -1)}{\text{ln}\left(\frac{|\text{csc}(a)+\text{cot}(a)|}{|\text{csc}(aS_{\infty})+\text{cot}(aS_{\infty})|}\right)} \label{eqn:R0_3}
\end{align}

For Step \ref{step3}, we find the inverse of $f$.

\begin{align}
f(x) = \text{sin}(ax) \implies f^{-1}(x)=\frac{\text{arcsin}(x)}{a} \label{eqn:inverse_3}
\end{align}

For Step \ref{step4}, substituting the expression for $\frac{1}{R_0}$ (\ref{eqn:R0_3}) and $f^{-1}$ (\ref{eqn:inverse_3}) into the overshoot equation ($Overshoot = f^{-1}\left(\frac{1}{R_0}
\right)-S_{\infty}$) yields:

\begin{align}
Overshoot &= \frac{1}{a}\text{arcsin}\left(\frac{a(S_{\infty}-1)}{\text{ln}\left(\frac{|\text{csc}(a)+\text{cot}(a)|}{|\text{csc}(aS_{\infty})+\text{cot}(aS_{\infty})|}\right)}\right) - S_{\infty} \label{eqn:overshoot_final3}
\end{align}

For Step \ref{step5}, differentiation of both sides with respect to $S_{\infty}$ and setting the equation to zero to solve for the critical $S_{\infty}^*$ yields:

\begin{align}
0 &= \frac{1}{a}\left(\frac{1}{\sqrt{1-\left(\frac{a(S_{\infty}^*-1)}{\text{ln}\left(\frac{|\text{csc}(a)+\text{cot}(a)|}{|\text{csc}(aS_{\infty}^*)+\text{cot}(aS_{\infty}^*)|}\right)}\right)^2}} \frac{\text{ln}\left(\frac{|\text{csc}(a)+\text{cot}(a)|}{|\text{csc}(aS_{\infty}^*)+\text{cot}(aS_{\infty}^*)|}\right)\cdot a - a(S_{\infty}^*-1)\cdot\frac{\frac{-|\text{csc}(a)+\text{cot}(a)|\cdot (-a \text{cot}(aS_{\infty}^*) \text{csc}(aS_{\infty}^*)-acsc^2(aS_{\infty}^*))}{(|\text{csc}(aS_{\infty}^*)+\text{cot}(aS_{\infty}^*)|)^2}}{\frac{|\text{csc}(a)+\text{cot}(a)|}{|\text{csc}(aS_{\infty}^*)+\text{cot}(aS_{\infty}^*)|}}}{\left(\text{ln}\left(\frac{|\text{csc}(a)+\text{cot}(a)|}{|\text{csc}(aS_{\infty}^*)+\text{cot}(aS_{\infty}^*)|}\right)\right)^2}\right)-1 \label{eqn:trans_3}
\end{align}
Solving this transcendental equation (\ref{eqn:trans_3}) requires first specifying the value of parameter $a$. For instance, specifying $a=1$ and solving the equation numerically yields $S_{\infty}^* = 0.1648...$ . 

For Step \ref{step6}, we use $S_{\infty}^*$ in the overshoot equation (\ref{eqn:overshoot_final3}) to obtain the value of the maximal overshoot for this model, $Overshoot^* |_{\substack {_{\beta (\text{sin}(aS))I}}}$. For $a=1$, we obtain:

\begin{align}
Overshoot^*|_{\substack {_{\beta (\text{sin}(S))I}}} &= 0.2931...
\end{align}
Thus, the maximal overshoot for incidence functions of the form $\beta (\text{sin}(S))I$ is 0.293... .

For Step \ref{step7}, we can calculate the corresponding $R_0^*$ using $S_{\infty}^*$ and (\ref{eqn:R0_3}).

\begin{align}
R_0^*|_{\substack {_{\beta (\text{sin}(S))I}}} &= 2.262...
\end{align}

\begin{figure}[ht]
\centering
\includegraphics[scale=0.2]{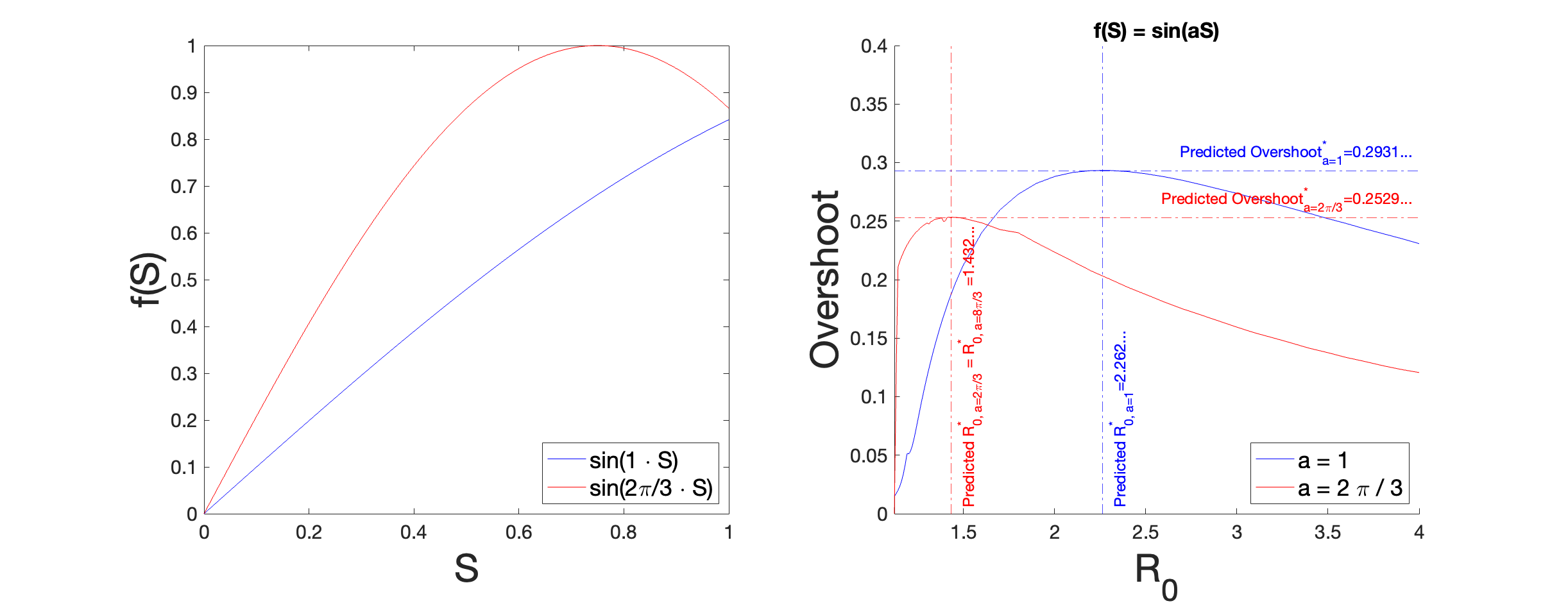}
\caption{The overshoot in a SIR model with periodic incidence. a) $f(S)=\text{sin}(aS)$ for different values of $a$. b) The overshoot as a function of $R_0$ for an SIR model with nonlinear incidence term of $\beta (\text{sin}(aS)) I$ for different values of $a$. The dashed horizontal lines for $Overshoot^*$ and the dashed vertical lines for $R_0^*$ are the theoretical predictions given by the calculations in the text. The solid curves are obtained from numerical simulations.} \label{fig:overshoot_sin}
\end{figure}

Now consider $a=\frac{2\pi}{3}$, which produces a non-monotonic $f(S)$ over the unit interval (Supplemental Figure \ref{fig:overshoot_sin}a). Solving (\ref{eqn:trans_3}) for $a=\frac{2\pi}{3}$ yields $S_{\infty}^* = 0.1163...$ . Repeating Steps \ref{step6} and \ref{step7} when $a=\frac{2\pi}{3}$ yields:

\begin{align}
Overshoot^*|_{\substack {_{\beta (\text{sin}(\frac{2\pi}{3}S))I}}} &= 0.2529... \\
R_0^*|_{\substack {_{\beta (\text{sin}(\frac{2\pi}{3}S))I}}} &= 1.432...
\end{align}

This leads to the question of what the applicable domain of $a$ is. The larger the value of $a$, the stronger the non-monotonicity of $f(S)$ is (Supplemental Figure \ref{fig:overshoot_sin}a). We can first eliminate values based on the condition $R_0 >\frac{1}{f(S)}$. Clearly that condition is violated if $f(S)$ is not positive, since $R_0$ is always positive because $\beta$ and $\gamma$ are both positive-definite. Since $f(S)=\text{sin}(aS)$, then $f(S)$ is negative when $a\in [\pi n,2\pi n], n=\mathbb{N}$. Furthermore, since we have a formula for $\frac{1}{R_0}$ using (\ref{eqn:R0_3}), we can set up the inequality explicitly.

\begin{align}
f(S=1)&>\frac{1}{R_0} \nonumber \\
\text{sin}(a\cdot1) &>  \frac{a(S_{\infty} -1)}{\text{ln}\left(\frac{|\text{csc}(a)+\text{cot}(a)|}{|\text{csc}(aS_{\infty})+\text{cot}(aS_{\infty})|}\right)} \nonumber
\end{align}

These result for the different cases of $f(S)=\text{sin}(aS)$ are shown in Supplemental Figure \ref{fig:overshoot_sin}b.

\subsubsection*{Time Series for Comparing Overshoot in Homogeneous and Heterogeneous Networks}
To provide intuition on why heterogeneous networks can have higher overshoot at larger transmission compared to their more homogeneous counterparts, we look in the time-series data of the epidemic.

\begin{figure}[ht]
\centering
\includegraphics[scale=0.7]{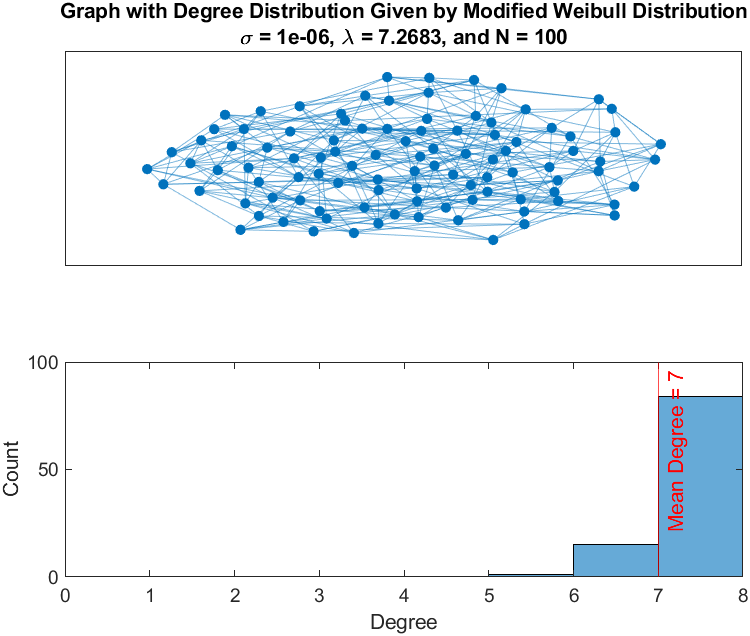}
\caption{Network representation and corresponding degree distribution for a network with $\sigma = 0.000001$ (very homogeneous). } \label{fig:homogeneousNetwork}
\end{figure}

Shown is an example from a network with a very homogeneous network distribution ($\sigma = 0.000001$, Supplemental Figure \ref{fig:homogeneousNetwork}) and a moderately heterogeneous network ($\sigma = 0.000001$, Supplemental Figure \ref{fig:heterogeneousNetwork}). The time series for a simulation on both networks in shown in Supplemental Figure \ref{fig:timeseriesNetwork}. Aside from $\sigma$, the other simulation parameters are held constant (i.e. number of nodes (N=100), transmission probability ($\tau=0.1$), recovery probability ($\rho=0.05$), number of initially infected individuals = 1, and $\langle k \rangle$= 7).

\begin{figure}[ht]
\centering
\includegraphics[scale=0.5]{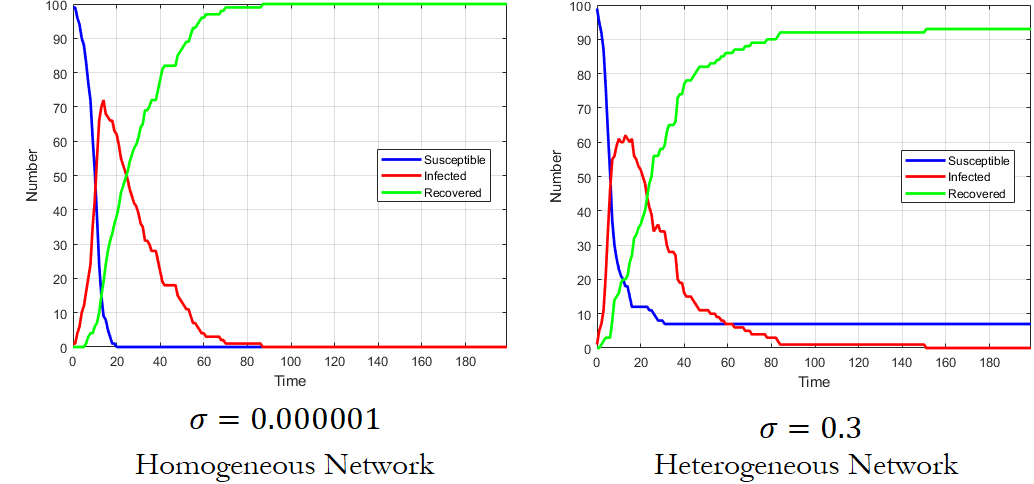}
\caption{Time series for two network simulations that vary only in their heterogeneity parameter ($\sigma$). Left) the network degree distribution is parameterized by $\sigma = 0.000001$ (very homogeneous). Right) The network degree distribution is parameterized by $\sigma = 0.3$ (moderate heterogeneity). For both networks, the other simulation parameters are $N=100, \tau=0.1, \rho=0.05$, number of initially infected individuals = 1, and $\langle k \rangle$ = 7.} \label{fig:timeseriesNetwork}
\end{figure}

\begin{figure}[ht]
\centering
\includegraphics[scale=0.7]{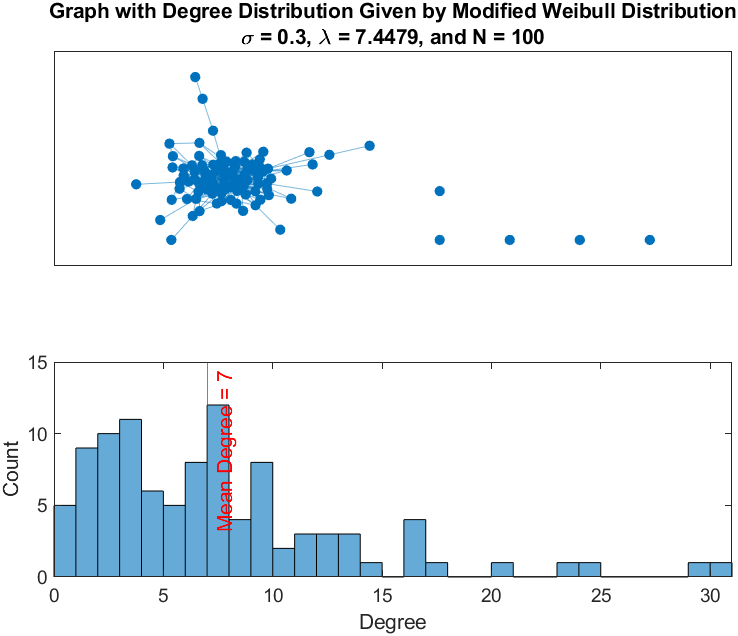}
\caption{Network representation and corresponding degree distribution for a network with $\sigma = 0.3$ (moderate heterogeneity).} \label{fig:heterogeneousNetwork}
\end{figure}

\begin{figure}[ht]
\centering
\includegraphics[scale=0.5]{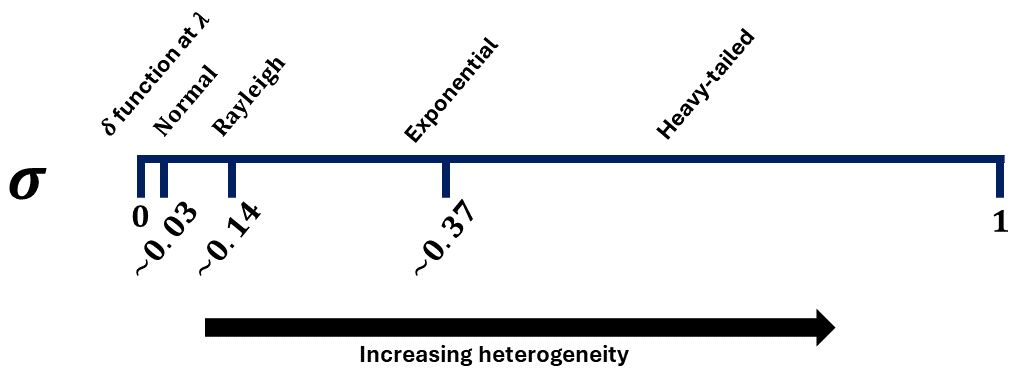}
\caption{Number line for the heterogeneity parameter ($\sigma$). Demarcated are common distribution shapes and their relative corresponding $\sigma$ value.} \label{fig:sigmaLine}
\end{figure}

\clearpage
\newpage
\begin{center}
\textbf{\large Supplemental Materials: Upper Bounds on Overshoot in the SIR Models with Nonlinear Incidence}
\end{center}
\subsection*{Code to Generate Figures}
Code executed in MATLAB R2023a.
\UseRawInputEncoding

\begin{lstlisting}[
frame=single,
numbers=left,
style=Matlab-Pyglike]
%%%% Run this first section as a script
numNodes = 200;
sigma = [0.000001, 0.00001, 0.0001, 0.001, 0.01, 0.05, 0.1, 0.2, 0.3, 0.5, 0.7];
weibullMean = 5;
transmissionProbability = (0:0.005:0.4);
recoveryProbability = 0.2;
initialInfected = 1;
time = 250;

numIterations = 150;

overshootVec = zeros(length(sigma), length(transmissionProbability), numIterations);
meanOvershootVec = zeros(length(sigma), length(transmissionProbability));
sigmaMatrix = zeros(length(sigma), length(transmissionProbability), numIterations);
tauMatrix = zeros(length(sigma), length(transmissionProbability), numIterations);

for x = 1:length(sigma)
    for y = 1:length(transmissionProbability)
        for z = 1:numIterations
            [epidemicDuration, finalAttackRate, overshoot] = nonlinearIncidenceScript(numNodes, sigma(x), weibullMean, transmissionProbability(y), recoveryProbability, initialInfected, time, centralParam, intervention, plotOption, tolThreshold);

            overshootVec(x,y,z) = overshoot;
            sigmaMatrix(x,y,z) = sigma(x);
            tauMatrix(x,y,z) = transmissionProbability(y);
        end
        meanOvershootVec(x,y) = mean(overshootVec(x,y,:));
    end
end

figure
hold on
custom_map = jet(length(sigma));
for s = 1:length(sigma)
    plot(transmissionProbability/recoveryProbability*weibullMean, meanOvershootVec(s,:),'-', 'Color', custom_map(s,:))
end

xlabel('Transmission Probability (\tau) \cdot Mean Degree / Recovery Probability (\gamma) ','FontSize',14)
ylabel('Overshoot','FontSize',14)
title(['N = ', num2str(numNodes), ', Mean Degree = ', num2str(weibullMean), ', \gamma =', num2str(recoveryProbability)],'FontSize',14)
legend(cellfun(@num2str, num2cell(sigma), 'UniformOutput', false),'FontSize',12)


%%%%%%%%%%%%%%%%%%%%%%%%%%%%%%%%%%%%%%%%%%%%%%%
function [overshoot] = nonlinearIncidenceScript(numNodes, sigma, weibullMean, transmissionProbability, recoveryProbability, initialInfected, time)

close all

lambda = weibullMean ./ gamma(1+1./(-\text{log}(sigma))); %Scale parameter. Sets middle of the probability distribution
adjacencyMatrix = generateModWeibullGraph(numNodes, sigma, lambda);

[epidemicDuration, finalAttackRate, overshoot] = simulateSIR(adjacencyMatrix, transmissionProbability, recoveryProbability, initialInfected, time, sigma, lambda);
end


%%%%%%%%%%%%%%%%%%%%%%%%%%%%%%%%%%%%%%%%%%%%%%%
function [overshoot] = simulateSIR(adjacencyMatrix, transmissionProbability, recoveryProbability, initialInfected, time, sigma, lambda)
% Input:
% - adjacencyMatrix: The adjacency matrix representing the network.
% - transmissionProbability: Probability of transmission (infection) between connected nodes per timestep..
% - recoveryProbability: Probability of recovery per timestep.
% - initialInfected: Number of initially infected individuals.
% - time: Total simulation timesteps.

% Number of nodes in the network
numNodes = size(adjacencyMatrix, 1);

% Initial state
susceptible = ones(numNodes, time);
infected = zeros(numNodes, time);
recovered = zeros(numNodes, time);
     
% Randomly select initially infected nodes
initialInfectedNodes = randperm(numNodes, initialInfected);
infected(initialInfectedNodes,1) = 1;
susceptible(initialInfectedNodes,1) = 0;
recovered(:,1) = zeros(numNodes, 1);
    
%% Simulation loop
for timestep = 2:time

    %% Calculate new infections probabilistically
        
    potentialInfections = transmissionProbability * (adjacencyMatrix * infected(:,timestep-1)) .* susceptible(:,timestep-1);
    newInfections = double(rand(numNodes,1) < potentialInfections);

    %% Calculate recoveries probabilistically
    potentialRecoveries = recoveryProbability * infected(:,timestep-1); % Recovery only occurs on previously infected nodes (not those generated in current time step)
    newRecovereds = double(rand(numNodes,1) < potentialRecoveries);

    % Update states
    susceptible(:,timestep) = susceptible(:,timestep-1) - newInfections;
    infected(:,timestep) = infected(:,timestep-1) + newInfections - newRecovereds;
    recovered(:,timestep) = recovered(:,timestep-1) + newRecovereds;
        
    if sum(infected(:,timestep)) == 0 && sum(infected(:,timestep-1)) > 0
        epidemicDuration = timestep-1;
    end
end

peakTime = find(sum(infected,1) == max(sum(infected,1)),1);
overshoot = (sum(susceptible(:,peakTime)) - sum(susceptible(:,end)))/numNodes;
end



%%%%%%%%%%%%%%%%%%%%%%%%%%%%%%%%%%%%%%%%%%%%%%%
function modWeibullGraph = generateModWeibullGraph(numNodes, sigma, lambda)
% Input:
% - numNodes: Number of nodes in the graph.
% - sigma: The shape parameter for a 2-parameter modified Weibull distribution.
% - lambda: The median parameter for a 2-parameter modified Weibull distribution that centers the distribution.

%% Generate Degree Distribution based on modified 2-parameter Weibull distribution (Ozbay and Nguyen)
alpha = -\text{log}(sigma);

distDraws = wblrnd(lambda, alpha, numNodes, 1);
degreeDistribution = round(distDraws);

%% Implement Configuration Model
adjacencyMatrix = zeros(numNodes);
numStubs = sum(degreeDistribution);
links = numStubs ./ 2;
stubs = zeros(numStubs, 1); %List of node IDs
dum = 0;
for i = 1:numNodes
    stubs((dum+1):(dum+degreeDistribution(i))) = i;
    dum = dum+degreeDistribution(i);
end

if mod(numStubs,2) == 0
    link_counter = 0;
    unique_nodes = numel(unique(stubs));

    %Generate the network with neither self interactions nor multiple edges
    while link_counter < links && unique_nodes > 1 && sum(sum(adjacencyMatrix(unique(stubs), unique(stubs)))) < (unique_nodes^2 - unique_nodes)
        edge = randsample(numStubs, 2); %Sample 2 nodes from the stub list without replacement
        if stubs(edge(1)) ~= stubs(edge(2)) && adjacencyMatrix(stubs(edge(1)), stubs(edge(2))) == 0
            adjacencyMatrix(stubs(edge(1)), stubs(edge(2))) = 1;
            adjacencyMatrix(stubs(edge(2)), stubs(edge(1))) = 1;
            link_counter = link_counter + 1;

            %Delete the used stubs from the stub list and update the number of
            %available stubs num_stubs. The entries of the stub list are put to NaN
            %and then deleted to ensure the right stubs are deleted (deleting
            %the 1st stub directly will shrink the list and make the second
            %stub index invalid)!!! Also update the number of unique node IDs
            %in the stub list
            stubs(edge(1)) = NaN;
            stubs(edge(2)) = NaN;
            stubs(isnan(stubs)) = [];
            numStubs = numel(stubs);
            unique_nodes = numel(unique(stubs));
        end
    end

else
    % Delete a stub to get total number of edges to an even number
    stubs(randi([1, numStubs], 1)) = [];
    numStubs = numStubs-1;
    links = numStubs ./ 2;
    stubs = zeros(numStubs, 1); %List of node IDs
    dum = 0;
    for i = 1:numNodes
        stubs((dum+1):(dum+degreeDistribution(i))) = i;
        dum = dum+degreeDistribution(i);
    end

    link_counter = 0;
    unique_nodes = numel(unique(stubs));

    %Generate the network with neither self interactions nor multiple edges
    while link_counter < links && unique_nodes > 1 && sum(sum(adjacencyMatrix(unique(stubs), unique(stubs)))) < (unique_nodes^2 - unique_nodes)
        edge = randsample(numStubs, 2); %Sample 2 nodes from the stub list without replacement
        if stubs(edge(1)) ~= stubs(edge(2)) && adjacencyMatrix(stubs(edge(1)), stubs(edge(2))) == 0
            adjacencyMatrix(stubs(edge(1)), stubs(edge(2))) = 1;
            adjacencyMatrix(stubs(edge(2)), stubs(edge(1))) = 1;
            link_counter = link_counter + 1;

            stubs(edge(1)) = NaN;
            stubs(edge(2)) = NaN;
            stubs(isnan(stubs)) = [];
            numStubs = numel(stubs);
            unique_nodes = numel(unique(stubs));
        end
    end

end

modWeibullGraph = adjacencyMatrix;
end


\end{lstlisting}
\end{document}